\documentclass[preprint,12pt]{elsarticle}

\usepackage{graphics}
\usepackage{graphicx}
\usepackage{epstopdf}
\journal{Chemical Physics}

\begin{document}

\begin{frontmatter}

\title{Infrared Laser Driven Double Proton Transfer. An Optimal Control Theory Study}
\author[bs,ur]{Mahmoud Abdel-Latif}
\author[ur]{Oliver K\"uhn\corref{cor1}}
\address[bs]{Chemistry Department, Faculty of Science, Beni-Suef University, Beni-Suef, Egypt}
\address[ur]{Institut f\"{u}r Physik, Universit\"{a}t Rostock, D-18051 Rostock, Germany}

\cortext[cor1]{Email: oliver.kuehn@uni-rostock.de }

\begin{abstract}
Laser control of ultrafast double proton transfer is investigated for a two-dimensional model system describing stepwise and concerted transfer pathways. The pulse design has been done by employing optimal control theory in combination with the multiconfiguration time-dependent Hartree wave packet propagation. The obtained laser fields  correspond to multiple pump-dump pulse sequences. Special emphasis is paid to the relative importance of stepwise and concerted transfer pathways for the driven wave packet and its dependence on the parameters of the model Hamiltonian as well as on the propagation time. While stepwise transfer is dominating in all cases considered, for high barrier systems concerted transfer proceeding via tunneling can make a contribution.
\end{abstract} 

\end{frontmatter}
\section{Introduction}
\label{sec:intro}
%
Laser pulse design has developed into a powerful method for controlling  molecular dynamics \cite{SfbBook,rice01,shapiro03,brixner03:418,wohlleben05:850,vivie-riedle07:5082}. In particular laser control of proton transfer reactions in the electronic ground state has triggered considerable theoretical efforts \cite{kuhn06:71}. A straightforward method for laser-driven proton transfer between two minima on a potential energy surface (PES) is the so-called pump-dump approach adapted from the study of isomerization reactions \cite{combariza91:10351,korolkov96:10874,jakubetz97:375}. In Ref. \cite{doslic98:292}  pump-dump control was demonstrated for a two-dimensional (2D) mo\-del mimicking the situation in malonaldehyde derivatives. The first pump-pulse promotes the system from the localized reactant state into a delocalized state above the reaction barrier from where it is dumped into the product well by a second pulse. The optimization of the pulse parameters can be done by hand, but it was found that optimal control theory (OCT) \cite{peirce88:4950,zhu98:1953,werschnik07:r175} applied to this problem essentially yields a similar pump-dump pulse field \cite{doslic98:9645,doslic99:1249}. OCT provides a means for controlling the pulse intensity via a penalty factor. Increasing the penalty for strong pulses in the case of proton transfer reactions resulted in a change of the mechanism from above barrier transfer to tunneling  through the barrier\cite{doslic98:9645,doslic99:1249}. In this case the resulting driving field resembles a few-cycle pulse and the associated dynamics in isolated and dissipative systems was investigated in some detail, e.g., in Refs. \cite{naundorf99:163,geva02:1629} (for applications of few-cycle pulses to isomerization reactions, see e.g. Refs. \cite{uiberacker04:11532,mitric07:031405}).  Besides these efforts along a quantum mechanical description, there have been attempts to trigger transfer in a double minimum PES using classical trajectory-based local control theory \cite{grafe06:271}.

Laser control of double proton transfer (DPT) reactions has received less attention. Nishikawa et al. \cite{nishikawa05:665} considered control of stepwise DPT in an asymmetrically substituted tetraflouro-porphyrin model using the stimulated Raman adiabatic passage (STIRAP) method. Assuming a stepwise transfer the description has been reduced to two independent one-dimensional potentials obtained from intrinsic reaction coordinate calculations. It was shown that STIRAP can achieve population inversion related to stepwise DPT on a time scale of some tens of picoseconds. Shapiro and coworker \cite{thanopulos05:14434} have investigated DPT in the context of DNA radiation damage and repair for a dinucleotide model. They developed a two-dimensional potential energy surface for DPT comprising a linear reaction coordinate for the concerted DPT supplemented by an out-of-plane squeezing type vibrational mode.  Employing coherently controlled adiabatic passage, detection and repair of DPT related mutation has been demonstrated.  Finally, Thanopulos et al. \cite{thanopulos08:445202} have shown the implications of  DPT control for single molecule charge transfer. Considering a thio-functionalized porphyrin deri\-va\-ti\-ve attached to four gold electrodes it was found from molecular orbital analysis that the  different trans-forms essentially provide orthogonal pathways for electron transfer. Laser controlled switching was discussed in a stepwise model comprising two bond coordinates for the first step and a single linear reaction coordinate for the second one. 

The motivation for the present study has been twofold. First, porphyrin-derivatives are not only interesting for charge transfer, but are used in the context of excitation energy transfer as well \cite{balzani03}. The rate for excitation energy transfer between two chromophores depends on the relative orientation between their transition dipole moments \cite{may04}. Since this orientation is affected by DPT in the electronic ground state one can, at least in principle, envision a laser-triggered ultrafast switch built into an energy transfer device.  Second, previous studies on controlled DPT considered either stepwise or concerted pathways. The decision for a pathway which is dominating in an actual system is often not that clear-cut (see discussion of the porphycene, a structural isomer of porphine in Ref. \cite{smedarchina07:314}) and may depend on the experimental conditions such as temperature \cite{lopez-del-amo09:2193}. Furthermore, the mechanism of DPT is commonly inferred from thermal rate constants or line splittings in absorption spectra. Driving a system with a control pulse, however, is likely to establish a pronounced nonequilibrium situation. Here, it is not evident that the system will follow the same reaction path as in the quasi-equilibrium case. 

In the following Section \ref{sec:model} we will present a two-di\-men\-sio\-nal model Hamiltonian which comprises concerted and stepwise DPT. This allows us to identify their relative importance in dependence on the applied control field. Laser control is achieved employing OCT in the frame of a multiconfiguration time-dependent Hartree (MCTDH) simulation of the wave packet dynamics \cite{beck00:1,MCTDH09}. Respective results will be discussed in Section \ref{sec:res}. 

%
\section{Theory}
\label{sec:model}
\subsection{Model Hamiltonian}
%
The DPT Hamiltonian will first be given in terms of single proton transfer coordinates $x_1$ and $x_2$ which are assumed to describe the linear translocation of the particles between donor and acceptor sites. The simplest form is obtained by combining two quartic potentials coupled via a bilinear interaction 
\begin{equation}
\label{eq:U12}
U_{\rm sym}(x_1,x_2) =\frac{U_0}{x_0^4}   \left[ (x_1^2-x_0^2)^2 + (x_2^2-x_0^2)^2 \right] - \frac{g U_0}{x_0^2} x_1x_2 \,.
\end{equation}
This type of potential has been used extensively by Sme\-dar\-chi\-na and coworkers \cite{smedarchina07:174513,smedarchina08:1291}. It has two equivalent glo\-bal minima (called "trans" in the following) and two equivalent local minimum (called "cis") which correspond to the transfer of a single proton. In eq. (\ref{eq:U12}), $U_0$ is the barrier for uncorrelated single proton transfer, $x_0$ is half the transfer distance, and $g$ is the dimensionless bilinear coupling strength. Since we are aiming at an analysis in terms of concerted and stepwise transfer, this potential is more conveniently expressed in terms of the symmetric and asymmetric transfer coordinates $x_s=(x_1+x_2)/2$ and $x_a=(x_1-x_2)/2$, respectively. This transformation gives for eq. (\ref{eq:U12})
\begin{eqnarray}
\label{eq:U0as}
U_{\rm sym}(x_s,x_a) & = & 2U_0 + \frac{U_0}{x_0^2} 
\left[(g-4)x_a^2-(g+4)x_s^2\right] \nonumber \\
 & + & \frac{2U_0}{x_0^4}(x_s^4+x_a^4+6x_s^2x_a^2) \,.
\end{eqnarray}
For the purpose of laser control of DPT it is useful to consider the case of asymmetric molecules in order to allow for a clear identification of initial and final states. Two types of asymmetry can be introduced into the model Hamiltonian, that is, with respect to the trans and cis states:
\begin{equation}
\label{eq:Uasym}
U_{\rm asym}(x_s,x_a)=\frac{\alpha_{\rm trans} U_0}{x_0} x_s + \frac{\alpha_{\rm cis} U_0}{x_0}x_a 
\end{equation}
where $\alpha_{\rm trans}$ and $\alpha_{\rm cis}$ are dimensionless parameters characterizing the detuning between the trans and cis states, respectively.

\begin{figure}
\begin{center}
\resizebox{0.7\columnwidth}{!}{
\includegraphics{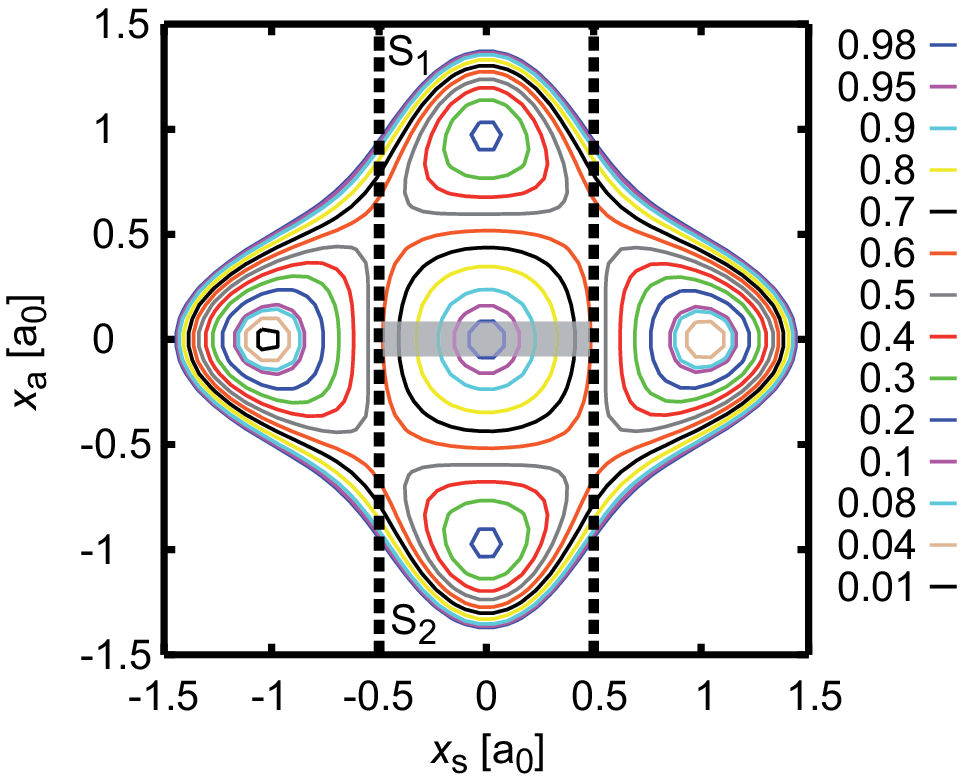}
}
\end{center}
\caption{(color online) Two-dimensional PES for the high barrier case ($U_0=2000$ cm$^{-1}$, $x_0=1$a$_0$, $g=0.2$, $\alpha_{\rm trans}=0.005$, and $\alpha_{\rm cis}=0.0$). The energy is given in units of the barrier height for concerted transfer (4428 cm$^{-1}$). The energetic difference between the two trans minima is 41 cm$^{-1}$. The two dashed lines mark the region which is passed during stepwise DPT. This doesn't include the shaded area whose width in $x_a$ direction corresponds to the width of the vibrational ground state and which is assigned to concerted DPT. Upper and lower stepwise DPT is marked by $S_1$ and $S_2$, respectively.  In the low barrier case (not shown) $U_0=800$ cm$^{-1}$, giving a barrier height of 1773 cm$^{-1}$. The barriers for stepwise DPT are 2430 cm$^{-1}$ and 975 cm$^{-1}$ and the energies of the cis minima are 821 cm$^{-1}$ and 330 cm$^{-1}$ for the high and low barrier case, respectively.}
\label{fig:pes}     
\end{figure}

We will choose parameters such as to highlight in particular cases of high and low barriers for the concerted DPT. The reference case will be the high barrier scenario with equivalent cis minima shown in Fig. \ref{fig:pes} (for parameters see figure caption).  Note that we do not address any specific system in our study, although the energetics of the PES can be considered to be typical for DPT molecules. Here, in particular porphycene derivatives offer a tunability of the energetics over a wide range (see, e.g., \cite{waluk07:245}).

The total Hamiltonian is given by
\begin{equation}
\label{eq:ham}
H=-\frac{\hbar^2}{2 m}\left( \frac{\partial^2}{\partial x_s^2}+\frac{\partial^2}{\partial x_a^2}\right) + U_{\rm sym}(x_s,x_a) + U_{\rm asym}(x_s,x_a)
\end{equation}
where it is assumed that the moving particles are H-atoms and thus the mass for the collective coordinates is $m=2 m_{\rm H}$. 

For the interaction with the laser field we assume that the permanent dipole moment depends  only linearly on the coordinates, i.e.,
\begin{equation}
\label{eq:field}
H_{\rm field}(t)= -(d_a x_a + d_s x_s ) E(t)
\end{equation}
where $d_{a/s}$ is the derivative of the dipole moment with respect to $x_{a/s}$ and $E(t)$ is the laser field. For the case of symmetric systems like porphycene the dipole moment changes only along the asymmetric coordinate (for simplicity we have chosen $d_a=1.0 $e). This will be used in the reference case. In order to investigate the principal effect of a dipole gradient along $x_s$ due to asymmetric substitution we will also consider a variation along $x_s$.

\subsection{Quantum Dynamics}
The two-dimensional time-dependent Schr\"odinger equation
\begin{equation}
\label{eq:sgl}
i\hbar \frac{\partial}{\partial t} \Psi(x_a,x_s;t) = (H+H_{\rm field}(t)) \Psi(x_a,x_s;t)
\end{equation}
has been solved using the MCTDH approach as implemented in the Heidelberg program package \cite{mctdh83}.
 To this end the two-dimensional wave function $\Psi(x_a,x_s;t)$ is represented on a grid in terms of a harmonic oscillator discrete variable representation (64 points within [-2.5:2.5]a$_{\rm 0}$).  The actual propagation is performed using the variable mean field  scheme  in combination with  a 6-th order  Adams-Bashforth-Moulton integrator. Using 20 single particle functions per coordinate the largest natural orbital populations have been typically on the order of $10^{-6}$. Selected eigenstates of the time-independent Hamiltonian, $\phi_i$, are obtained by improved relaxation \cite{meyer06:179}.
 
The shape of the laser field is determined using OCT, that is, the following functional is maximized
at some final time $T$ \cite{rabitz03:64,sundermann1896:00}
\begin{equation}
\label{eq:J}
J(E, T) = \langle \Psi(T) | \hat{O} | \Psi(T) \rangle - \kappa \int_0^T dt \frac{E^2(t)}{\sin^2(\pi t /T)} 
\end{equation}
where $\hat{O}$ is the target operator and $\kappa$ the penalty factor. The iterative optimization of the field has been performed by using the implementation within the MCTDH program package by Brown and coworkers as detailed in Ref. \cite{schroder08:850}. Note that their approach doesn't follow the standard iteration procedure which is not suitable in the context of MCTDH. Instead an extrapolation scheme is used following the suggestion of Ref. \cite{wang06:014102}.

The pulses will be characterized in terms of their XFROG trace
 \begin{equation}
I_{\rm XFROG}(\omega,t)=\left|\int d\tau E(\tau)G(\tau-t)e^{-i\omega \tau}\right|^2
\end{equation}
where $G(t)$ is a step-like gate function with Gaussian tails \cite{schroder08:850}.

It will be assumed that initially the system is in the vibrational ground state which is localized in the left part of the potential due to the asymmetry term $\alpha_{\rm trans}$. The target operator will be taken as the projector onto that state which is localized in the right well of the potential, i.e. $ \hat{O}=|\phi_1\rangle\langle \phi_1|$ (see Fig. \ref{fig:3}). In order to elucidate the mechanism of DPT, i.e., stepwise vs. concerted, we have defined step-like operators dividing the different regions of transfer between the two trans-forms as shown in Fig. \ref{fig:pes}. Here, transfer is counted as concerted if the wave packet passes a narrow range in the vicinity of the second order saddle point of the PES. The width of that range is, of course, arbitrary and we have used the width of the ground state distribution. Furthermore, note that this definition of ''stepwise«« doesn't imply the existence of a stable intermediate in the sense of traditional kinetic considerations.

The results reported below have been obtained for different numbers of  OCT iterations starting with a sin$^2$-shaped guess field.  The convergence of the  control functional has been monotonic. The iteration number has been chosen such that the change in the control yield between two iterations was below 10$^{-4}$.  The guess field frequency we have chosen to correspond, for the high barrier case, to the transition between the ground state and the first excited state along the $x_s$ coordinate which is slightly red-shifted as compared with the strong IR active first excited state along the $x_a$ coordinate. For the low barrier case the resonance to the $x_a$ coordinate has been chosen. Lowering the guess frequency, e.g. by 20 cm$^{-1}$, resulted in zero yield, an increase by the same amount gave a similar convergence behavior of the control yield.
%
\section{Results and Discussion}
\label{sec:res}
%
\subsection{Dependence on Pulse Duration}
\begin{figure}
\begin{center}
\resizebox{1.0\columnwidth}{!}{
\includegraphics{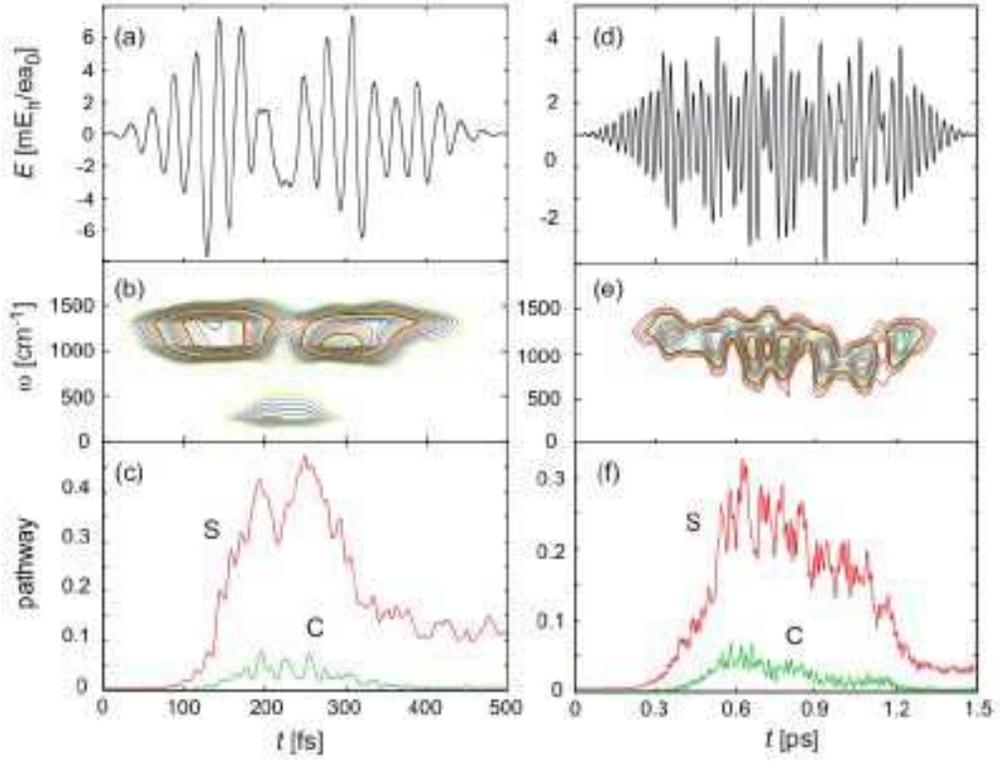}
}
\end{center}
\caption{(color online) Laser pulses obtained from OCT using (a) $T=500$ fs (guess field $E_0=7$ mE$_{\rm h}$/ea$_{\rm 0}$) after 291 iterations, (d) $T=1.5$ ps, (guess field $E_0=0.95$ mE$_{\rm h}$/ea$_{\rm 0}$) after 165 iterations ($\kappa=1.5$ a.u., $\omega=1194$ cm$^{-1}$). Panels (b) and (e) show the respective XFROG traces for a gate function of width 1000 a.u.  and Gaussian tails of width 1000 a.u. (contours (b) 2.6 to 8.9 in steps of 0.3 a.u., (e) 1.4 to 4.6 in steps of 0.2 a.u. ). Panels (c) and (f) give the probability of being in the C and S regions of the PES.}
\label{fig:2}     
\end{figure}

The OCT optimization problem depends on the final time $T$ at which the optimization goal shall be reached. The obvious question is to what extent will the OCT field and the associated wave packet dynamics depend on the pulse duration.  Fig. \ref{fig:2} compares OCT pulses for $T=500$ (a) and 1500 fs (d). Inspecting their XFROG traces in panels (b) and (e) we notice that upon increasing $T$  one obtains a pulse train like structure with an increasing number of overlapping subpulses.

\begin{figure}
\begin{center}
\resizebox{0.85\columnwidth}{!}{
\includegraphics{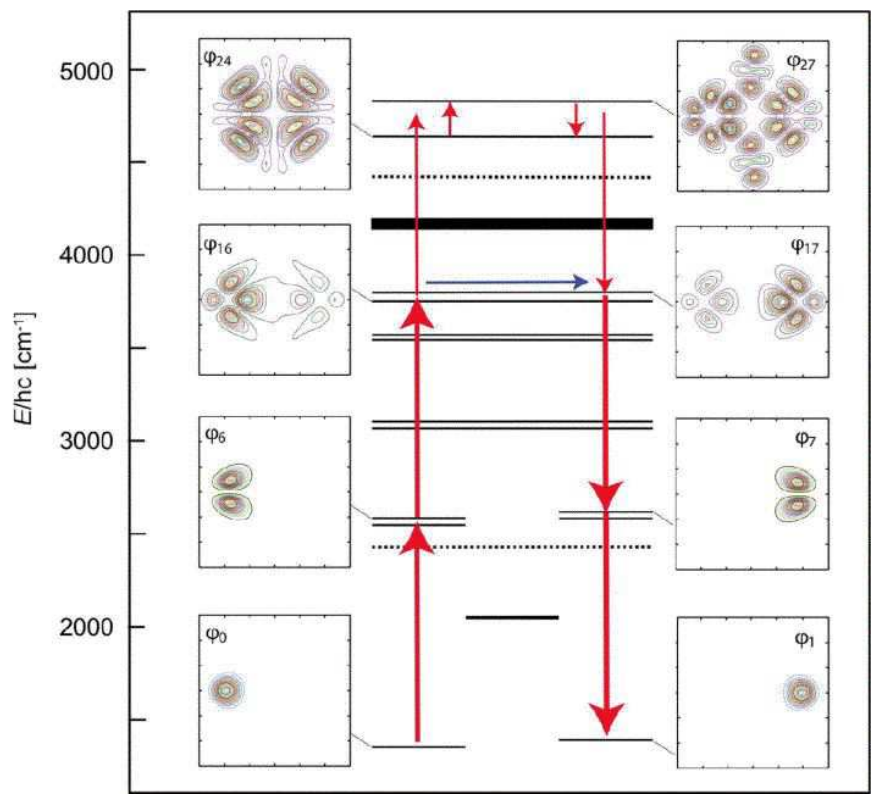}
}
\end{center}
\caption{(color online) Level scheme for the high barrier case and densities of selected eigenstates which get appreciably populated
during the laser driven transfer. The lengths of the vertical solid bars indicate the range of delocalization of the state with respect to the symmetric coordinate $x_s$ (e.g. $\phi_0/\phi_1$ are localized in the left/right trans minimum, $\phi_2/\phi_3$ ($\sim$2000 cm$^{-1}$) around $x_s=0$, i.e. in the cis minimum, etc.). The dashed lines correspond to the energy of the barriers for concerted and stepwise transfer. The density plots cover the range $[-1.5:1.5]a_0$ along the vertical $x_a$ and the horizontal $x_s$ axes. The vertical arrows indicate the major pathway for laser driven transfer, the horizontal one shows the pathway for concerted transfer. (Note that around 4200 cm$^{-1}$ there are several close lying states.)}
\label{fig:3}     
\end{figure}

\begin{figure}
\begin{center}
\resizebox{0.7\columnwidth}{!}{
\includegraphics{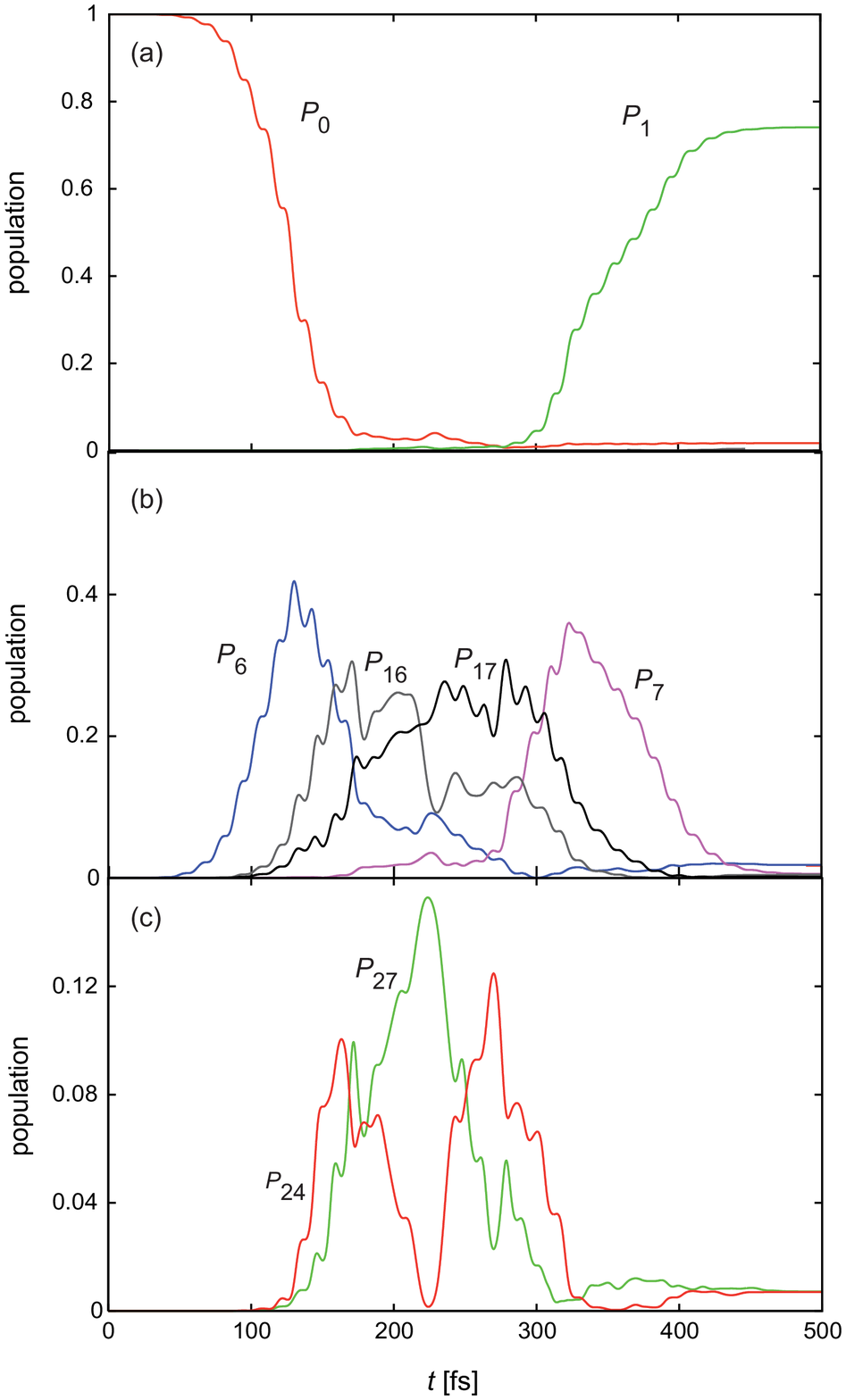}
}
\end{center}
\caption{(color online) Population dynamics for the pulse of Fig. \ref{fig:2}a and the states shown in Fig. \ref{fig:3}.}
\label{fig:pophigh500}     
\end{figure}

The dynamics can be analyzed in terms of populations of eigenstates of the field-free Hamiltonian. A level scheme as well as densities for selected states are given in Fig. \ref{fig:3}. In total there are 24 states below the barrier for concerted DPT. Also shown are the localized initial, $\phi_0$, and target, $ \phi_1$, states which differ by 38 cm$^{-1}$ for the chosen $\alpha_{\rm trans}$. The population dynamics for $T=500$ fs is presented in Fig. \ref{fig:pophigh500}. Panel (a) shows that the initial state is almost completely depopulated during the first $\sim$180 fs whereas the target state becomes populated after $\sim$300 fs reaching a final population of 0.74. Since the dipole moment changes along the $x_a$ direction only,  we would expect that the laser pulse excites a wave packet along this direction. Indeed the initial excitation is to state $\phi_6$ which has a node along $x_a$. But already the next state which is populated, i.e. $\phi_{16}$, is of mixed character containing excitations along both directions. This mixing is, of course, a consequence of  the anharmonicity of the potential. Both states are excited by the first pulse within about 200 fs. The mismatch between the $\phi_0\rightarrow \phi_6$ and $\phi_6 \rightarrow \phi_{16}$ transitions amounts to a  65 cm$^{-1}$ redshift which is covered by the pulse spectrum. The latter  also shows a slight down-chirp. Due to the delocalized nature of states $\phi_{16}$ and $\phi_{17}$ the first pulse actually excites a superposition of these two states, what can be seen from the rise of the population of state $\phi_{17}$ around 100 fs in Fig. \ref{fig:pophigh500}. The second pulse which is centered around 300 fs acts as a dump pulse transferring the populations according to the scheme $\phi_{17} \rightarrow \phi_7 \rightarrow \phi_1$.
In between the two main pulses there is a minor feature at $\sim$220 fs spectrally located around 200 cm$^{-1}$. It is related to the population of states which are energetically above the barrier for concerted transfer (see, Fig. \ref{fig:3}). In fact the two main pulses excite and de-excite state $\phi_{24}$ according to $\phi_{16} \rightarrow \phi_{24}$ and $\phi_{24} \rightarrow \phi_{17}$, respectively. The subpulse centered around 200 cm$^{-1}$ switches populations according to $\phi_{24} \rightarrow \phi_{27} \rightarrow \phi_{24}$. This effect is most likely not relevant for the DPT control.

\begin{figure}
\begin{center}
\resizebox{0.7\columnwidth}{!}{
\includegraphics{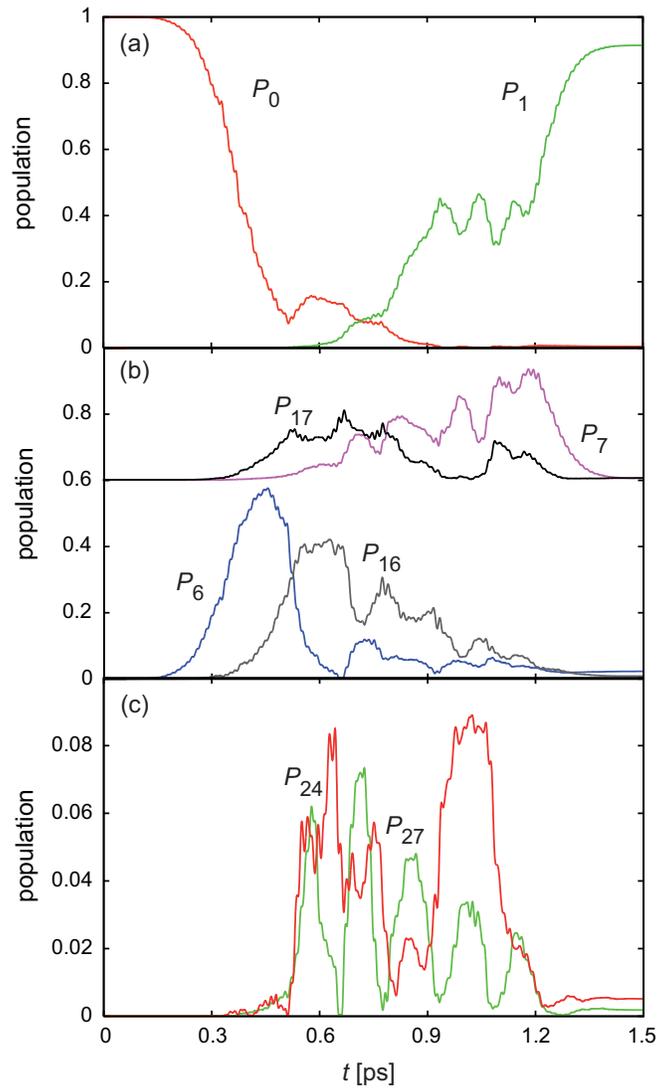}
}
\end{center}
\caption{(color online) Population dynamics for the pulse of Fig. \ref{fig:2}d and the states shown in Fig. \ref{fig:3}. ( In panel (b) $P_7$and $P_{17}$ have been offset by 0.6 for clarity.)}
\label{fig:pophigh1500}     
\end{figure}

The population dynamics for the $T=1500$ fs case is shown in Fig. \ref{fig:pophigh1500}. As compared with the $T=500$ fs case it is more complex. Only the effect of the first pulse is comparable, that is, the excitation scheme 
 $\phi_0\rightarrow \phi_6 \rightarrow \phi_{16}/\phi_{17}$ applies. Some features of the dynamics triggered by the other subpulses are: (i) The initial state is repopulated around 600 fs and subsequently depopulated until about 900 fs. (ii) A population/depopulation is also observed for the target state around 1000 fs. (iii) Above barrier states are populated directly by transitions from below barrier states. Finally, the population of the target state at 1500 fs is 0.91, thus exceeding the one for the 500 fs case, i.e. the convergence is faster for the longer pulse.

Since the reference case contains a dipole moment in $x_a$ direction only one expects that the laser driven wave packet  is bound to move toward the product well in a stepwise fashion, i.e. passing through the regions S$_1$ and S$_2$ in Fig. \ref{fig:pes}. Indeed this is the major pathway as shown for different pulse durations in Figs. \ref{fig:2}c and f. We further notice that the rise of the target state population is closely related to the decay of the probability of being in the regions S$_1$ and S$_2$. For example, for $T=1500$fs (Fig. \ref{fig:2}d-f) the latter has decayed by $\sim$ 1200 fs which coincides with a steep rise of $P_1$. In other words, it takes about 1200 fs for the major part of the laser-driven wave packet to reach the product well via a stepwise mechanism. Similar arguments hold for the $T=500$ fs case where this transition takes place around 300 fs. 

Comparing the results for the two  propagation lengths one notices a difference as far as the dynamics in the range of concerted DPT is concerned. Inspecting the densities of the different populated states it is clear that the contribution to the concerted pathway is mostly due to the  pair $\phi_{16}$ and $\phi_{17}$. A superposition of these states is excited already by the first pulse and the subsequent dynamics will be that of a two-level system. The energy mismatch between these two states of 51 cm$^{-1}$ yields a time scale of $\sim$650 fs for a round trip between the two minima, i.e. the transfer from the reactant to the product well takes about 325 fs. Given the approximate nature of this estimate, e.g. due to preparation process, one finds this time scale for the motion through the concerted region in the $T=500$ fs case. For the $T=1500$ fs case the time that  the wave packet spends in the $C$ region is about 900 fs and from Fig. \ref{fig:2}f one notices that there are three maxima of the $C$ region population during this period. Since $\phi_{16}$ and $\phi_{17}$ are energetically below the barrier, one can conclude that the contribution to concerted DPT is mostly due to tunneling.  

%
\subsection{Dependence on Model Parameters}
\label{sec:para}
\begin{figure}
\begin{center}
\resizebox{0.7\columnwidth}{!}{
\includegraphics{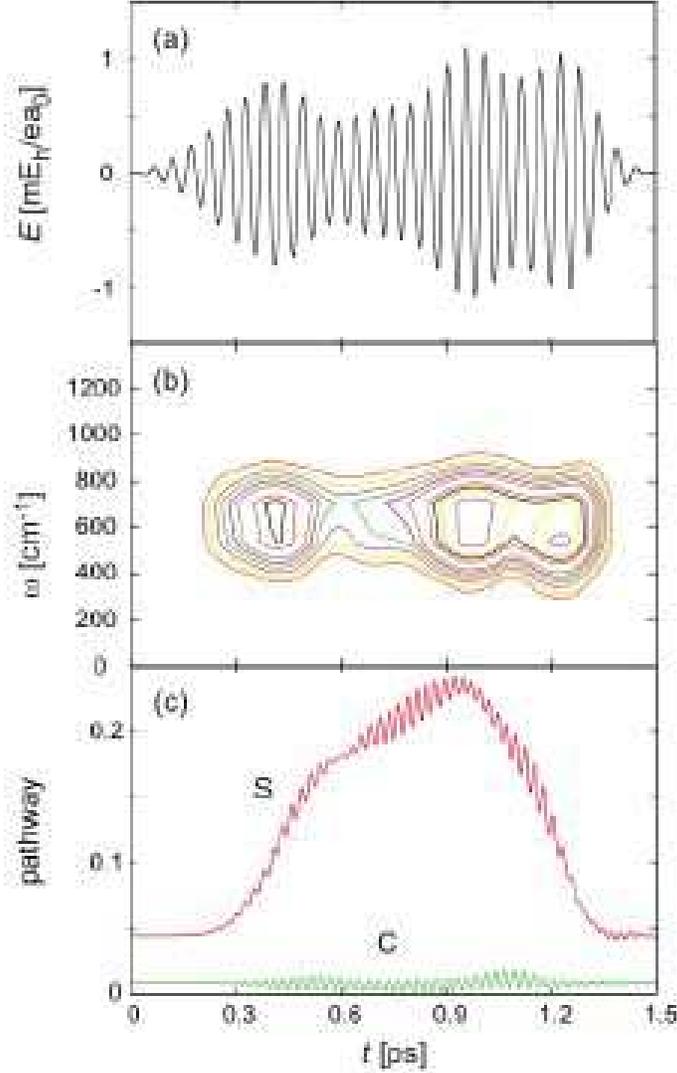}
}
\end{center}
\caption{OCT simulation for the low barrier case. (a) Optimized field for  $T=1500$ fs (guess field $E_0=0.95$ mE$_{\rm h}$/ea$_{\rm 0}$, $\omega=637$ cm$^{-1}$, $\kappa=1.5$ a.u.) after 60 iterations, (b) XFROG traces for a gate function of width 1000 a.u.  and Gaussian tails of width 1000 (contours from  0.5 to 1.7 in steps of 0.1), (c) probability of being in the two cis  (S) or in the concerted (C) transfer region.}
\label{fig:6}     
\end{figure}

\begin{figure}
\begin{center}
\resizebox{1.0\columnwidth}{!}{
\includegraphics{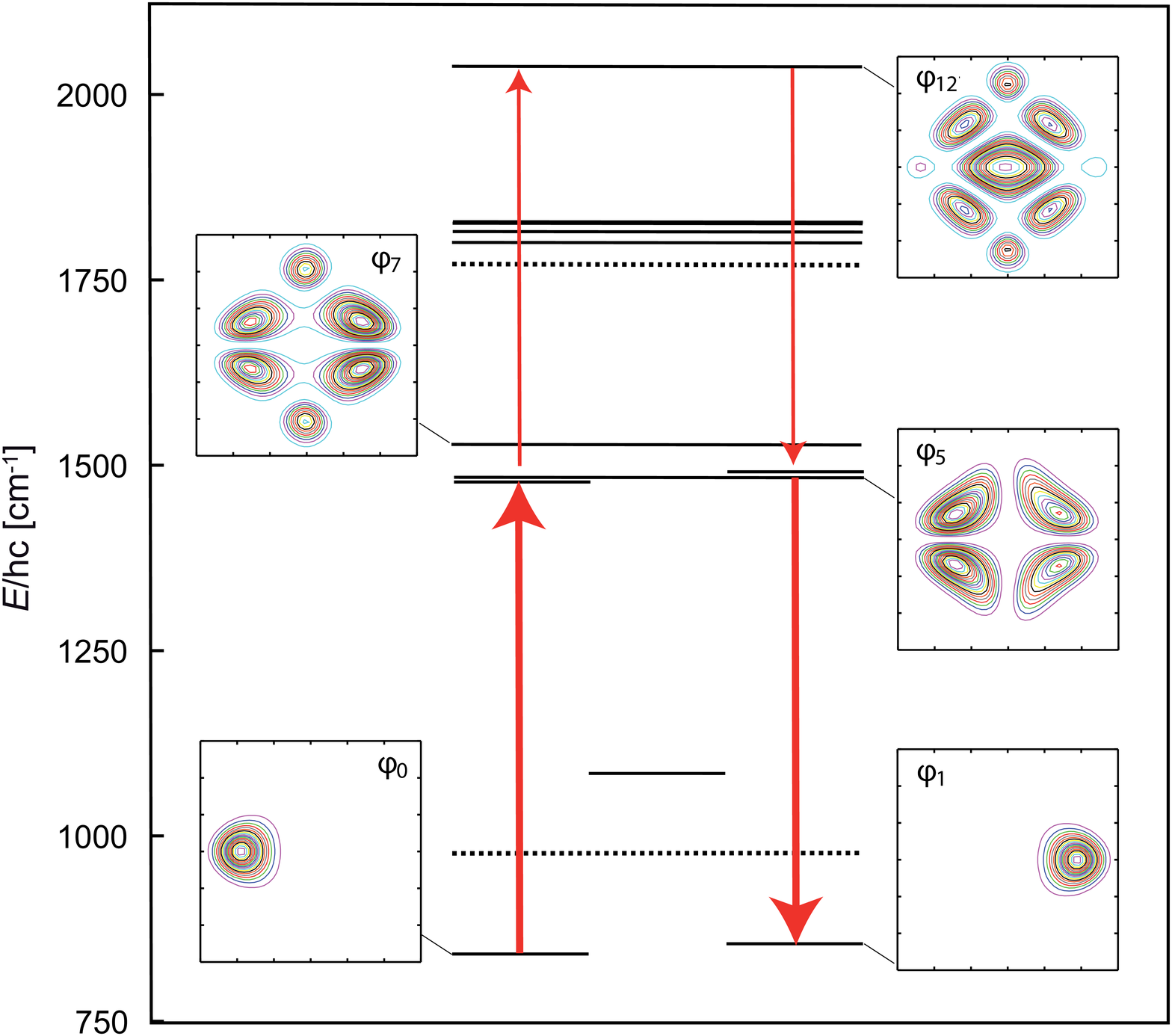}
}
\end{center}
\caption{(color online) Level scheme for the low barrier case and densities of selected eigenstates which get appreciably populated
during the laser driven transfer. The density plots cover the range $[-1.5:1.5]a_0$ along the vertical $x_a$ and the horizontal $x_s$ axes. (for meaning of arrows and vertical bars, see Fig. \ref{fig:3})}
\label{fig:7}     
\end{figure}

\begin{figure}
\begin{center}
\resizebox{0.8\columnwidth}{!}{
\includegraphics{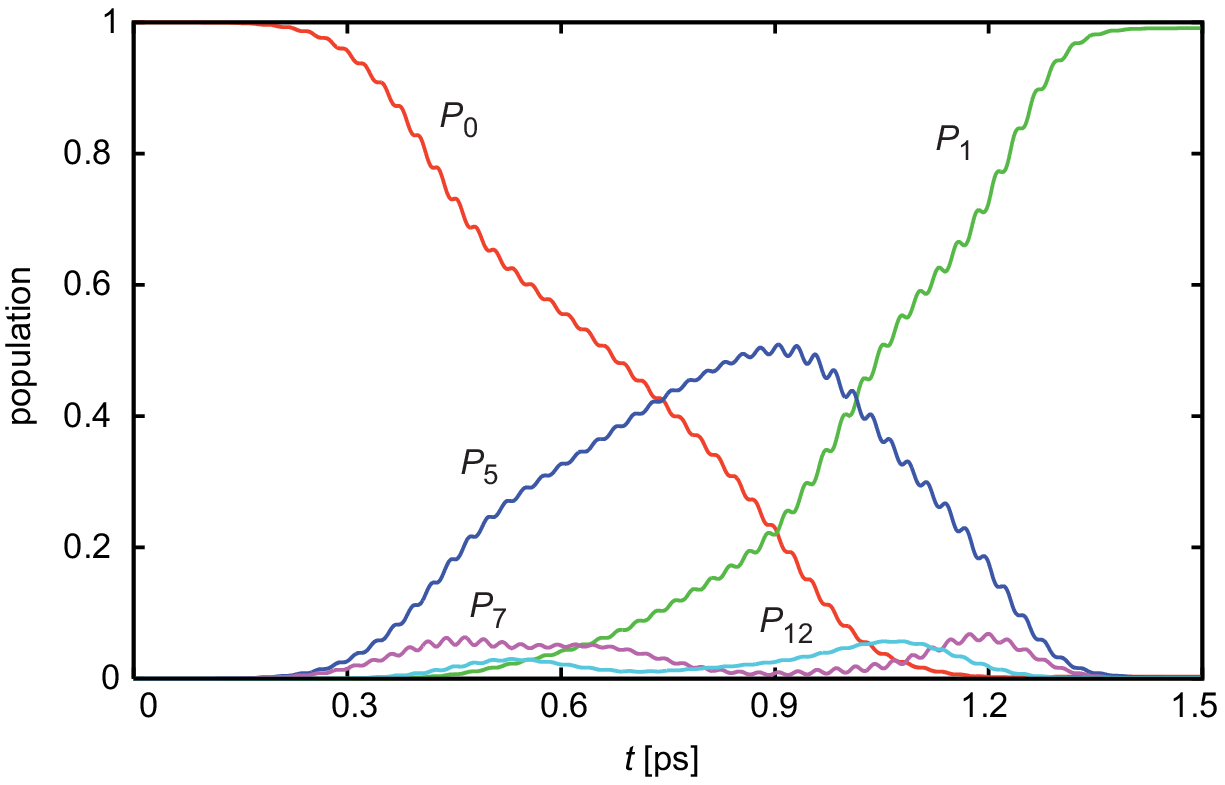}
}
\end{center}
\caption{(color online) Population dynamics for the pulse of Fig. \ref{fig:6}a and the states shown in Fig. \ref{fig:7}.}
\label{fig:8}     
\end{figure}

%
Lowering the reaction barriers has a substantial influence on the OCT field and the wave packet dynamics. Exemplarily we show the case of $T=1500$ fs in Fig. \ref{fig:6}. First, we notice that, in contrast to the high barrier case, the OCT pulse consists of only three subpulses. These pulses drive the wave packet essentially via the stepwise pathways as seen in Fig. \ref{fig:6}c. This is  somehow counter-intuitive since one would expect that upon lowering the barrier for concerted DPT, tunneling becomes even more effective. However, for the present parameters there is essentially only one state below the barrier having locally an excited $x_a$ character, namely $\phi_5$; see Fig. \ref{fig:7}. This state is not appreciably mixed with other states, say of $x_s$ excitation character. Instead it is delocalized encompassing both trans as well as the cis regions. The population dynamics  in Fig. \ref{fig:8} shows that the dominant pathway is indeed $\phi_0 \rightarrow \phi_5 \rightarrow \phi_1$ triggered in a pump-dump like fashion. The spectrally broad pulses also excite state $\phi_7$ which in turn causes an excitation of an above barrier state, $\phi_{12}$. The latter state clearly will lead to a contribution of the concerted pathway, but this time not by virtue of tunneling. However, it's effect is rather minor as can be traced from the population of the concerted region in Fig. \ref{fig:6}c. Finally, we note that due to the sparser energy level structure the OCT iteration converges faster and a 99\% population of the target state is achieved.

\begin{figure}
\begin{center}
\resizebox{0.7\columnwidth}{!}{
\includegraphics{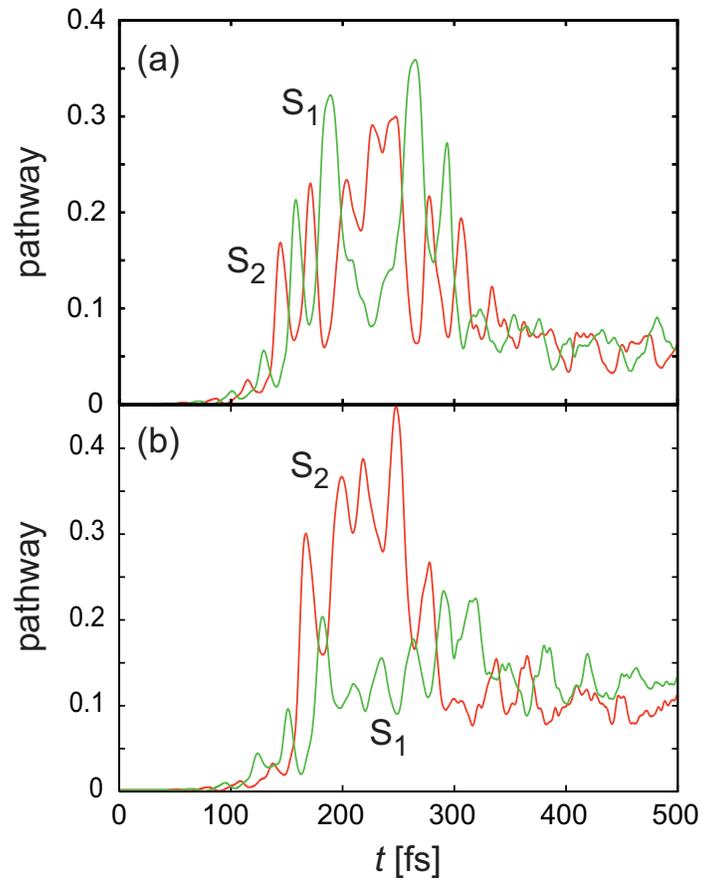}
}
\end{center}
\caption{(color online) Comparison of the populations of the two cis regions (see, Fig. \ref{fig:pes}) for the optimized $T=500$ fs pulse in the high barrier case for (a) equal ($\alpha_{\rm cis}=0$) and (b) different ($\alpha_{\rm cis}=0.01$) energies of the cis minima. }
\label{fig:9} 
\end{figure}

Next we explore the effect of adding an asymmetry to the two cis configurations for the high barrier reference case. We have chosen $\alpha_{\rm cis}=0.01$ which translates into an energetic  difference of 77 cm$^{-1}$ between the two cis minima. In the symmetric case the populations of the different cis regions S$_1$ and S$_2$ reflect the oscillation of the wave packet excited in the more stable trans well as seen in Fig. \ref{fig:9}a. That means, if the major part of the wave packet moves via S$_1$ there is a minimum in the population of S$_2$ and vice versa. Overall there is  no net preference for a pathway. If the S$_2$ region is energetically more favorable OCT prefers this way as seen from Fig. \ref{fig:9}b. 

Finally, we address the case where there is a nonvanishing dipole moment along both directions, exemplarily we have chosen $d_a/d_s=4$. Here, direct excitation of the $x_s$ coordinate becomes possible and one would expect the importance of the concerted pathway to increase. Indeed this is the case and for the optimized field the integrated populations for both pathways during the time interval up to 1500 fs are obtained to give a ratio of C$_{\rm tot}/{\rm S_{tot}}=0.144$. This value exceeds the  0.130 obtained from Fig. \ref{fig:2}f. However, the extra gain in concerted DPT is small. In fact the OCT pulse populates a state of $x_s$ excitation character ($\phi_4$) which is $\sim$36 cm$^{-1}$ below $\phi_6$. The population, however, doesn't exceed 10\% and is subsequently also promoted to $\phi_{16}$. Thus the smallness of the gain in concerted DPT is due to the  fact that $\phi_{16}$, which plays the dominant role for the concerted DPT, can be reached from the $x_a$ and $x_s$ fundament excitation state $\phi_6$ and $\phi_4$, respectively. That is, there is no substantial net effect due to the (competing) excitation of the symmetric coordinate.

In summary, optimal control theory has been used to devise laser fields which trigger ultrafast double proton transfer in a simple two-dimensional potential supporting stepwise and concerted mechanisms. The optimized pulses are of (multiple) pump-dump character.  stepwise transfer was found to be dominant for all cases considered. Concerted transfer has been shown to become relevant if almost degenerate pairs of states exist which are mostly localized in the reactant or product well but can be excited simultaneously. In order to excite such a superposition state for the situation where the dipole moment changes along the asymmetric coordinate only, the states have to be of mixed character with respect to symmetric and asymmetric excitations.  This situation is likely to be met for high barrier cases which support many locally excited states, that attain mixed character with increasing energy due to the anharmonicity of the potential energy surface. In such a situation a dipole change along the symmetric coordinate does not provide an additional means for concerted DPT.
For low reaction barriers already fundamental excitations of the asymmetric coordinate can be delocalized such as to provide a direct pump-dump pathway involving a mostly single intermediate state.  

The present use of a rather simple model Hamiltonian facilitates application  to specific molecular systems by virtue of its straightforward parametrization. An extension of the latter might be necessary, however, since it is known that the effect of heavy atom motions on the proton transfer and the mechanism of DPT can be substantial. In this respect the used OCT-MCTDH approach is ideally suited for studying driven multidimensional quantum dynamics.
\section*{Acknowledgment}
This work has been financially supported by a scholarship  from the Ministry of Higher Education of the Arab Republic of Egypt. We are grateful to M. Schr\"oder (University of Alberta) for his helpful comments concerning the MCTDH-OCT calculations.
\clearpage\newpage
%
\clearpage\newpage
%
\newpage
%
\clearpage\newpage
%
\clearpage\newpage
%
%
%

\end{document}